# Strain effects on the wear rate of severely deformed copper

Evander Ramos[a], Takahiro Masuda[b], Sina Shahrezaei[c], Zenji Horita[b], Suveen Mathaudhu[ac]*

[a] Department of Mechanical Engineering, University of California, Riverside, United States
[b] Department of Materials Science and Engineering, Kyushu University, Fukuoka 819-0395, Japan
[c] Department of Materials Science and Engineering, University of California, Riverside, United States

* Corresponding author: smathaudhu@engr.ucr.edu

**Abstract**
A variety of severe plastic deformation (SPD) techniques have been developed to process materials to high strains and impart microstructural refinement. High pressure torsion (HPT) is one technique that imparts inhomogeneous strain to process discs with low strain in the center and high strain at the outer edge. In the literature, this inhomogeneity is typically ignored when characterizing wear properties after HPT. In this work, the wear rate of pure copper discs processed by HPT was characterized by conducting dry sliding reciprocating wear tests at a few judicious locations on the discs. From only two discs, the wear resistance across many ranges of strains was captured. These measurements agreed with the literature for other SPD processes at varying strains. Wear rates dropped and plateaued at about 25% that of the unprocessed state when processing past equivalent strains of around 15, after which microstructural and microhardness saturation has also been observed. Some indication of a relationship between the direction of the imposed SPD shearing and the sliding wear direction was also observed. The incremental microstructure, microhardness, and wear resistance evolution past equivalent strains of ~15 indicate that for high purity copper these properties receive no clear benefit from higher SPD strains.

## 1. Introduction

Severe plastic deformation (SPD) techniques have been well-documented for processing materials to obtain ultrafine-grained (UFG) or nanograined microstructures [1]. Among SPD processes, high-pressure torsion (HPT) is attractive because the microstructural refinement it imparts is quick and efficient although also inhomogeneous throughout the disc at low equivalent strains (i.e. number of turns) [1]. This inhomogeneity varies due to the amount of imposed equivalent strain ($\varepsilon$) throughout the disc during processing, which is commonly calculated from equation (1), where $r$ = radial distance, $N$ = number of rotations, and $t$ = thickness [1].

$$\varepsilon = \frac{2\pi r N}{\sqrt{3}\, t} \qquad (1)$$

In the literature, hardness measurements are commonly used to indicate the degree of microstructural homogeneity throughout HPT discs [2]. However, other properties such as wear rate or friction can also vary with the equivalent strain, and they may do so in ways that do not follow the same trends indicated by hardness variations throughout discs.

Inhomogeneity is often ignored in reports of the wear response of HPT materials. Continuous ball-on-disc wear testing (ASTM G99) has been used [3–6], but these tests only cover localized and distinct levels of equivalent strain (i.e. sliding along one specific radial distance), and do not characterize the wear behavior in other regions across the diameter of the discs. Other studies [7–16] have used linear reciprocating ball-on-flat wear tests (ASTM G133) which cover a wider range of strain regimes and microstructures. But, these reciprocating wear tests are typically conducted at only one location without comparing to other locations on the discs. Thus, reported wear rates for HPT materials have been generalizations based on tests of one section of a disc. Furthermore, a review article on the wear performance of SPD-processed UFG materials has highlighted conflicting reports of both improvements and reductions in wear rates observed after HPT for different material systems, a phenomenon that was also seen in some materials after equal channel angular extrusion (ECAE) [17]. This article hypothesized that in some cases the wear resistance increased due to the smaller grains and higher hardness from SPD, whereas in other cases the wear resistance decreased due to reduction in ductility and strain hardening capability after SPD processing. It is also possible that the specific regions of equivalent strain covered by the wear tests can influence the resulting wear response measured, and so experiments analyzing different locations of processed samples can be contributing to these conflicting reports. Thus, the accuracy of wear response characterization in HPT materials can be improved by accounting for the variation in equivalent strain throughout discs, facilitating observations of correlations between these properties and the microstructures produced.

To probe wear properties with respect to the equivalent strain from HPT processing, linear reciprocating wear tests were conducted at judicious locations covering different ranges of strains. These results were compared to the literature to observe trends with equivalent strain across various processing methods. Pure copper was chosen as the material system due to the wealth of SPD literature accumulated for comparison, along with its relevance for commercial applications. The present work reports that wear rate follows trends in microstructural and microhardness evolution with processing strain, but also a relationship between wear path and strain path has been observed.

## 2. Materials and Methods
High purity copper (99.99 wt%) with an initial grain size of ~20 µm was sliced into discs 10 mm in diameter and 1 mm thick before being annealed at 873K for 1 hour. These discs were subjected to HPT in air and at room temperature at a pressure of 6 GPa for 1 and 10 rotations. After HPT, the flashing was removed, and surfaces were ground flat up to 1200 grit. The discs were kept in air at room temperature for about 2 years before characterization. To prepare the surfaces for characterization, the discs were electropolished at 4.7 V for 90 s in a 200 mL solution of 10:5:5:1 $H_2O$:phosphate:ethanol:IPA with 1 g urea [18]. Immediately afterwards, the discs were rinsed thoroughly in deionized water followed by IPA. Vickers microhardness was measured with a Phase II Vickers Microhardness Indenter with a 200 g load and 15 s dwell time. Indentations were taken along four diameters and spaced 0.5 mm apart. Microstructural characterization was carried out using an FEI Nano NovaSEM 450 scanning electron microscope (SEM) with a concentric backscatter detector (CBS). Average grain size was determined using Abrams three-circle procedure. A foil was extracted from the half radius of the 10 rotation sample using an FEI

Quanta™ 3D 200i focused ion beam (FIB). This foil was used for transmission electron microscopy (TEM) using a Titan Themis 300 at 300kV.

Unlubricated linear reciprocating ball-on-flat wear tests were conducted on full uncut discs using a Nanovea mechanical tester in accordance with ASTM G133. A normal force of 5 N was applied for a stroke length of 3 mm at a frequency of 0.25 s$^{-1}$ for a total sliding distance of 10 m and a total time of under 2 hours. Tests were conducted in air at a relative humidity of 44±2% and at room temperature (22±1.5°C) for all samples. The wear test parameters used in this study were chosen to ensure mild sliding and to be similar to other wear studies on pure copper to facilitate comparison. Mild sliding was achieved by using slow sliding speed to avoid excessive frictional heating and a low load with hard and tough wear balls to limit external transfer onto the copper surface. A new 3 mm diameter ball of 6% Co-cemented WC was used for each track and inspection determined that negligible ball wear occurred with little adherence of wear debris. These sliding parameters help to isolate the mechanical deformation of the ultrafine grained copper microstructure during wear. A foil from the center of the high strain tangential wear track on the 10 rotation disc was extracted using FIB lift-out technique and imaged using a Tecnai12 TEM at 150kV. This wear track was chosen for an initial exploration of the subsurface microstructural evolution due to it having been deformed to the highest strain level.

On each disc, wear tests were carried out at four locations, one of which spanned along the radius between 1 and 4 mm away from the center (herein called "radial tests"). The other three tests were centered at various distances from the center such that they would be parallel to the tangent of the disc (herein called "tangential tests"). The test locations were deliberately chosen so the tangential tests were near the highest, lowest, and average equivalent strain regions of the radial tests. In other words, the tangential tests covered small ranges of strains corresponding to those encountered at the center and either end of the radial test. Only one test was conducted at each range of strains, so no standard deviations could be provided for the HPT processed disc tests. This is in part due to dimensional limitations of the processed discs, which may be a contributing factor into why such tests have not been conducted on HPT discs in the literature. The resulting wear tracks were cleaned with compressed air and evaluated using a Nanovea optical profilometer and analysis software to determine the profile of the scar with 10 µm resolution parallel to the track and 1 µm resolution perpendicular to the track. Specific wear rate was calculated by dividing the measured wear volume at the conclusion of the test by the total sliding distance and load.

## 3. Results
### 3.1. Typical microstructures after HPT
As in prior works on HPT Cu, the microstructures varied radially throughout the processed discs [19–23]. Micrographs of the characteristic microstructures near the center and edge of discs after 1 and 10 rotations of HPT are shown in Figure 1. In the 1 rotation sample (Figure 1A), the grain sizes were heterogeneously distributed with a bimodal combination of large grains and small sections of highly refined grains. The center of the disc had the largest average grain size, while the outer edge had more regions of highly refined grains. From the SEM images taken at various locations throughout the disc, the overall average grain size was 700 nm ± 200nm. Conversely, the 10 rotation sample had mostly highly-refined grains with some large grains scattered throughout the disc, giving an average grain size of 290 nm ± 150nm. These large grains were most prevalent near the center and gradually became scarcer toward the edges.

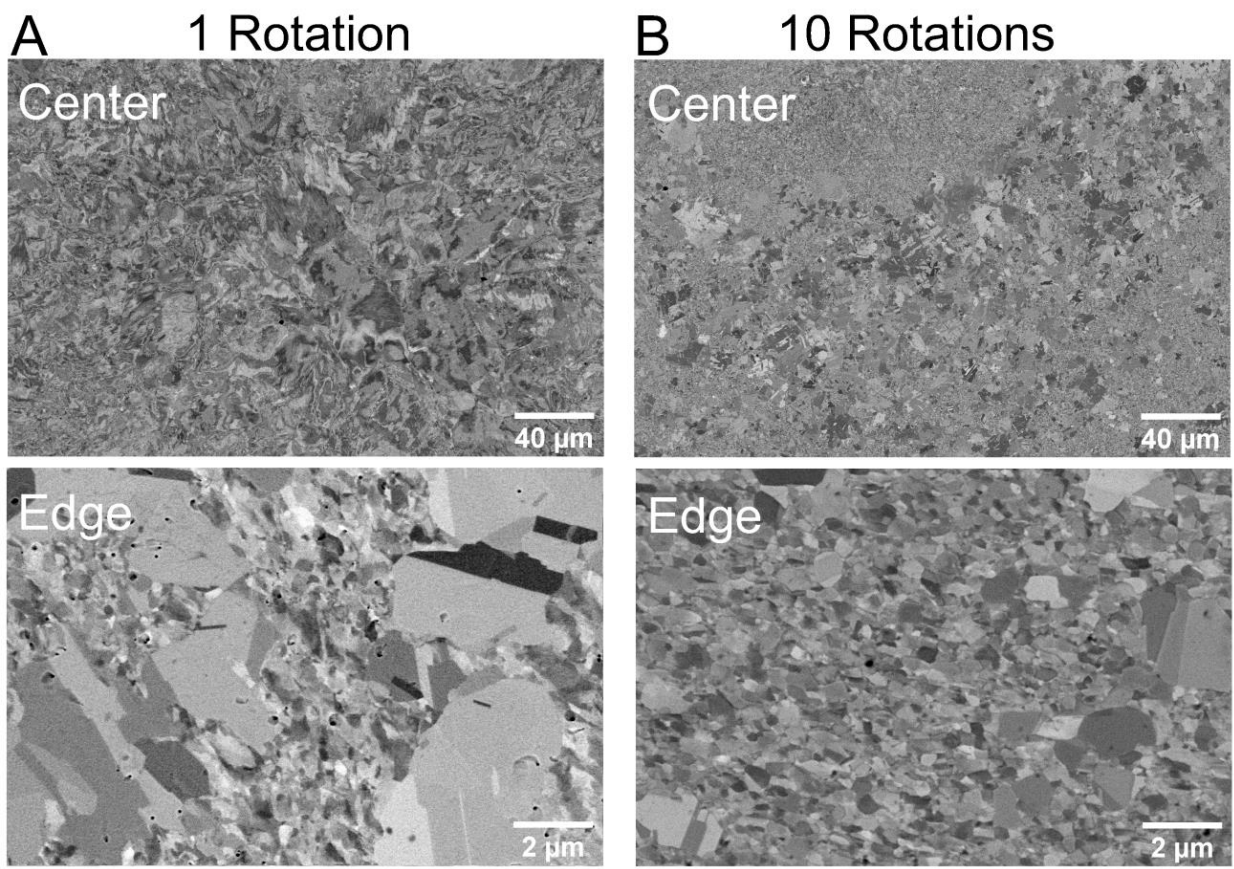

Figure 1: Micrographs displaying representative microstructures near the center and the edge of copper processed by HPT after 1 (A) and 10 (B) rotations.

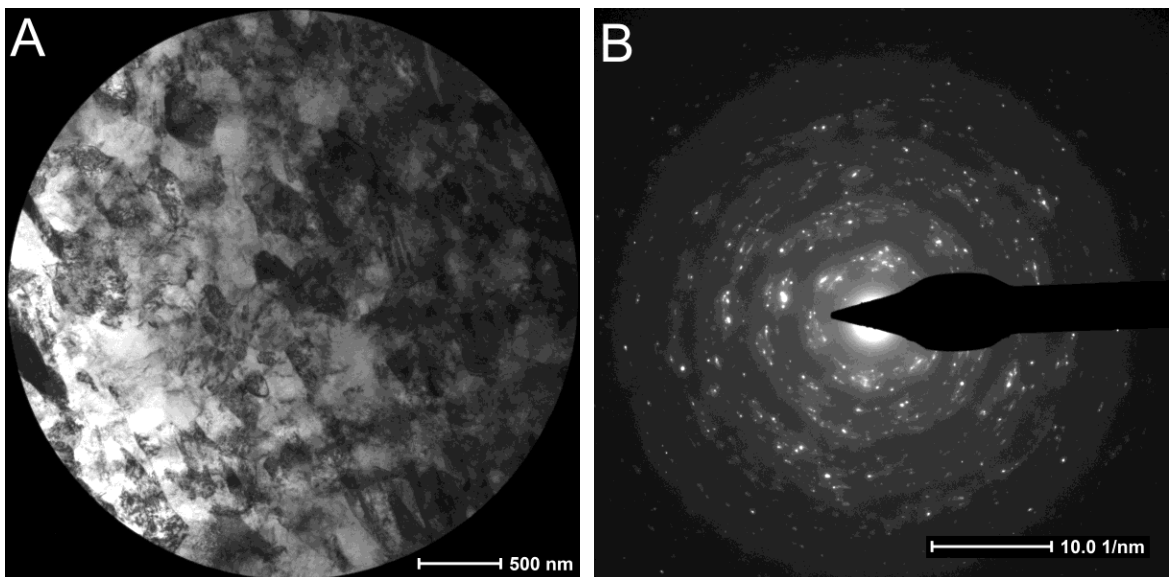

Figure 2: Transmission electron micrograph of a foil taken from the half radius of the 10 rotation sample (A) and the accompanying SAED pattern for this location (B).

The microstructure at the mid-radius of the 10 rotation sample seen using TEM is shown in Figure 2A, further validating the previous observations from SEM. The SAED pattern of this location shown in Figure 2B exhibits a diffuse ring pattern, indicating the presence of many different grain orientations as well as a high degree of lattice strain. The average grain size of 250 nm ± 66 nm measured from TEM images agreed with the averages determined from SEM.

*3.2. Microhardness inhomogeneity after 1 rotation, saturation after 10 rotations*
The microhardness results for the current study are shown in Figure 3A. Microhardness saturated at around 130 HV for the outer section of the 10 rotation sample, with a slight decrease near the center. For the 1 rotation sample, microhardness values are highest near the center, decreasing dramatically to 80 HV at 1.5 mm from the center, after which it steadily increases with radial distance. The error bars shown represent one standard deviation but are too small to be seen behind the points themselves for most of the data. The largest error is seen in the central portion of the 1 rotation sample, where the values decreased precipitously.

*3.3. Higher strain generally decreases wear rate*
Wear results for the four wear tests on each disc are shown in Figure 3B. The inset schematic shows the locations of the reciprocating tests on the HPT discs. For the unprocessed disc, the average specific wear rate for the four tests was $7.0 \times 10^{-5} \pm 1.3 \times 10^{-5}$ mm$^3$/N-m. With only one test done for each location on the processed discs, it is difficult to tell the statistical significance of the differences between each tested location. Nevertheless, the specific wear rates for all tested locations on the processed discs were between $2.7 \times 10^{-5}$ and $1.0 \times 10^{-5}$ mm$^3$/N-m, clearly outside the range of the unprocessed disc. Comparing the two processed discs, each test from the softer 1 rotation disc had a higher wear rate compared to the corresponding test on the harder 10 rotation disc. This is in agreement with the Archard equation, stated in equation (2), which relates wear volume V, a dimensionless constant K, normal load N, total sliding distance L, and the hardness of the contacting surface H [24].

$$V = \frac{KLN}{H} \qquad (2)$$

The Archard equation can also explain differences between the three tangential tests on each disc. For the 10 rotation disc, microhardness increases slightly with equivalent strain, and accordingly the wear rate decreases from the low to high strain tests. For the 1 rotation disc, since the low strain test partially traverses the harder central region of the disc, it has a wear rate comparable to the high strain test, with the softer middle strain section having a higher wear rate. Interestingly, the wear rate values for the tangential high strain test on the 1 rotation disc and the tangential low strain test on the 10 rotation disc are similar despite the difference in measured microhardness between the two regions ($1.6 \times 10^{-5}$ mm$^3$/N-m and 900 MPa vs. $1.7 \times 10^{-5}$ mm$^3$/N-m and 1360 MPa, respectively). Most apparent is that the radial tests, which covered the largest ranges of strain across the discs, had higher wear rates compared to the tangential tests.

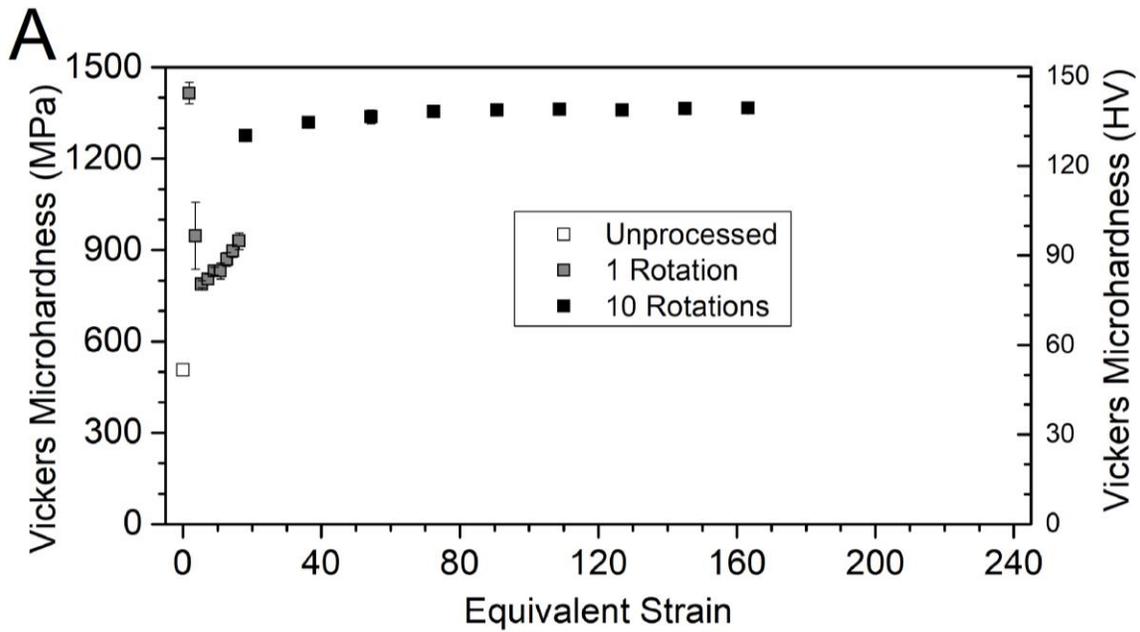

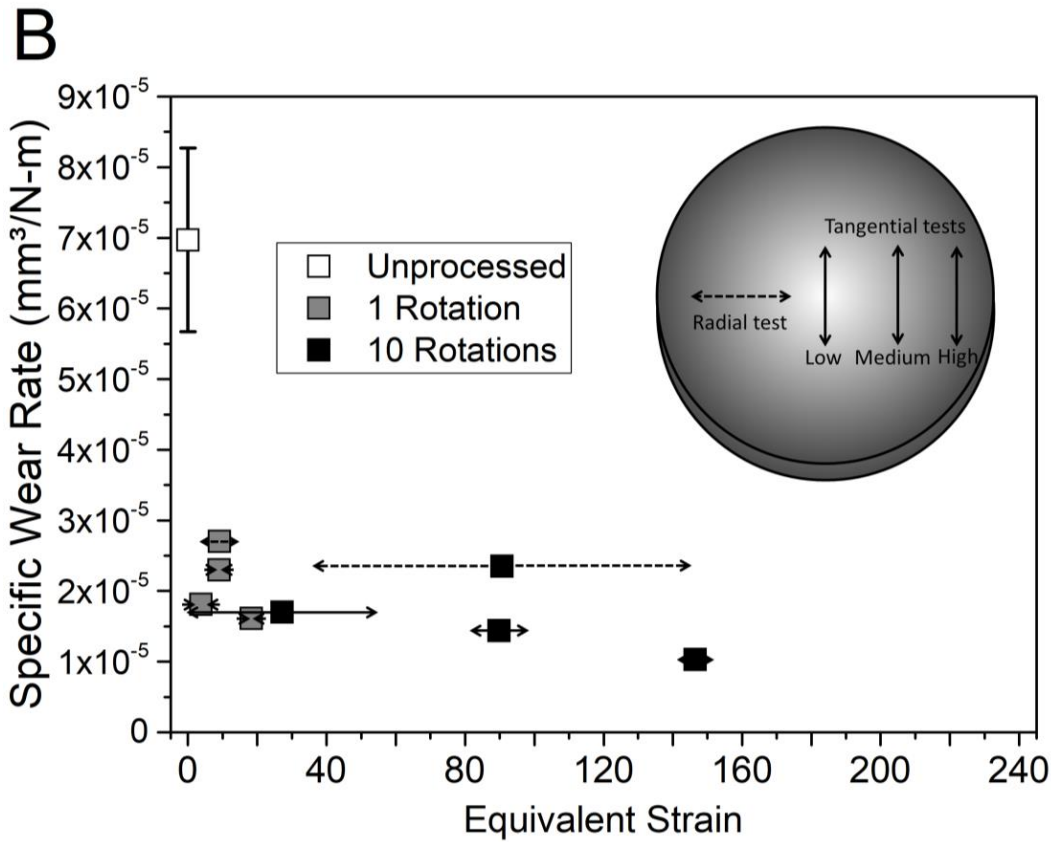

Figure 3: A) Variation in microhardness vs. equivalent strain throughout the discs after 1 and 10 rotations of HPT. B) Variation in specific wear rate vs. equivalent strain for the four tests done on each disc. The schematic inset on the right shows the locations of the reciprocating wear tests on the disc, with a gradient in color to indicate the evolution from low strain in the center (light) to high strain at the outer edge (dark). For the wear data, each data point represents the average equivalent strain along the wear track, while the x-spread represents the range of strains covered.

To compare the results from the current work to data from the literature for high purity copper processed by various SPD methods, the variation in normalized wear rate with equivalent strain was compiled as shown in Figure 4. Works that use various different wear test parameters can be compared via normalized wear rates, as has been previously shown for aluminum matrix composites [25]. Throughout this paper, normalized wear rate is calculated by the wear rate of the processed sample divided by the wear rate of the unprocessed control, both measured under the same conditions. Thus, a normalized wear rate of 1 indicates no change in wear due to processing, while a higher normalized wear rate indicates that the wear rate for the processed condition was higher than the unprocessed control group. A normalized wear rate less than one indicates an improvement in wear rate, e.g. a normalized wear rate of 0.5 indicates a 50% reduction in wear rate due to processing. Whenever a work used multiple wear testing parameters, the only results included are from tests with speed and load closest to the conditions used in the current work, namely 5 N load and 1.5 mm/s sliding speed.

The strain imposed by an SPD process is quantifiable by calculating equivalent strain. But, for some works in Figure 4, the processing imposes complex strain that is not as readily quantifiable or comparable, and those cases are simply left as bands across the equivalent strain axis. While not an SPD method, one work used electro-deposition to create samples approaching the critical grain size for the breakdown of the Hall-Petch effect [26]. Thus, it can be thought of as a limiting case for grain refinement encountered in SPD processes. The work using dynamic plastic deformation (DPD) is distinct from all the others in that it is the only one conducted in a cryogenic environment with high strain rates of $10^3$ s$^{-1}$ [27]. The deformation strain for this process has been shown to efficiently refine grains toward the nanoscale even at small strains, and thus obfuscates direct strain comparison with the other processes included in Figure 4 [28]. For the works using surface mechanical grinding treatment (SMGT) [29] and surface mechanical attrition treatment (SMAT) [30], strain decreases as a function of depth from the processed surface. The wear tests in the SMAT work were performed with varying loads from 5-60 N and saw that larger loads had higher normalized wear rates due to those tests penetrating toward the larger grained, less strained regions farther below the processed surface [30]. Another work used columnar microstructural features in a steel sample to determine SMAT strain, and from an exponential fit the near surface shear strain was estimated to be around 90 [31]. Thus, the data points for such surface processing techniques can be reasonably thought of as corresponding to high equivalent strains.

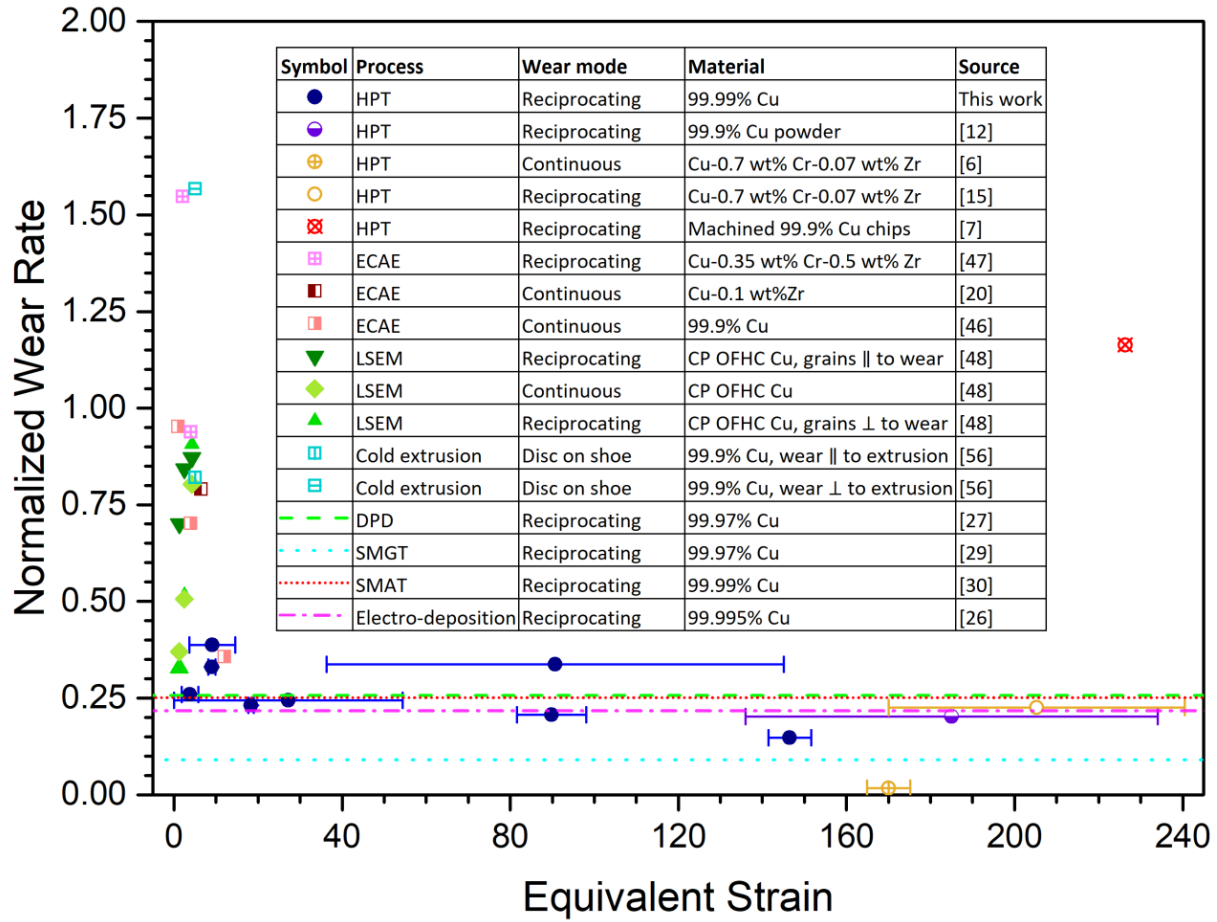

Figure 4. Normalized wear rate variation with equivalent strain for the wear tests from this work compared with other works on >99% pure copper processed by various methods. Normalized wear rate is calculated by dividing the reported wear rate or wear volume of the processed condition by the result for the unprocessed condition, thus each work is essentially normalized against itself.

*3.4. Friction response and microstructural evolution below the wear track is similar to prior works*
Although not the focus of the current work, frictional data was also measured continuously throughout each test. For each track, the coefficient of friction (COF) evolved rapidly, reaching an early peak after less than a few hundred centimeters of total sliding, after which it dropped slightly and thereafter slowly climbed to a steady state value for the last few meters of the test. The average steady state COF for the four tests was 0.59 ± 0.07, 0.57 ± 0.04, and 0.52 ± 0.04 for the unprocessed, 1 rotation, and 10 rotation discs, respectively. Unlike the wear rate results, the average steady state COF measurements for the radial tests were not clearly different from the tangential tests. There was higher variance with the COFs for the unprocessed disc, probably due to these wear tracks being bumpier and less smooth. These friction values are lower than some other reported values from the works in Figure 4, which may be attributed to the slower speeds used for this report. Another study that also used mild reciprocating sliding parameters on copper reported similar COF values [32]. Generally, the similarities in the friction measurements for all tests indicates that all tests are dominated by the same wear mode.

The microstructure below the wear track of the high strain tangential test on the 10 rotation disc is shown in Figure 5. A region with a nanocrystalline (NC) microstructure extending below the wear track to a depth of ~1.5 µm can be seen in Figure 5A. Below this NC region, the grains have an UFG structure similar to that which was present before the wear testing. Prior studies of copper microstructures after wear have also exhibited regions of nanocrystalline grains extending to various depths below the wear surface followed by a region of UFG microstructure [33,34]. In the current study, it is difficult to clearly differentiate between the wear affected UFG subsurface region typically formed in wear and the original UFG microstructure. The grains below the nanocrystalline region do appear slightly coarser than from before the wear testing (compare to Figure 2A). The nanocrystalline grains immediately below the surface are shown more clearly in Figure 5B. The SAED from this region shown in Figure 5C gives no indication that any WC from the sliding ball transferred onto the wear surface, consistent with the intended mild sliding condition and observations of the ball surface and mass after testing. A large amount of $Cu_2O$ is seen corresponding not only to the oxidized surface of the foil, but also to the oxides formed at the surface and integrated into the mechanically mixed layer. The presence of these oxides stabilizes the nanocrystalline structure near the wear surface, helping to maintain a stable average grain size on the order of 10 nm. The d-spacing for FCC cobalt (111) is 0.205 nm, similar to the d-spacing for Cu (111) which is 0.208 nm, so there is the possibility that a small amount of the cobalt binder in the ball could have transferred into the wear surface and also contributed to stabilizing the near surface nanocrystalline region. One last notable feature of Figure 5A is the high contrast, slightly angled crack just above the UFG layer. It may have originated within the subsurface due to the plastic deformation of the sliding or alternatively was formed at the wear surface and absorbed in the nanocrystalline layer, although it could also just be an artifact from the FIB lift out process. Further characterization of the subsurface microstructures formed below the other wear tracks should be done in future work to investigate such potential cracking in the wear affected regions of SPD processed materials.

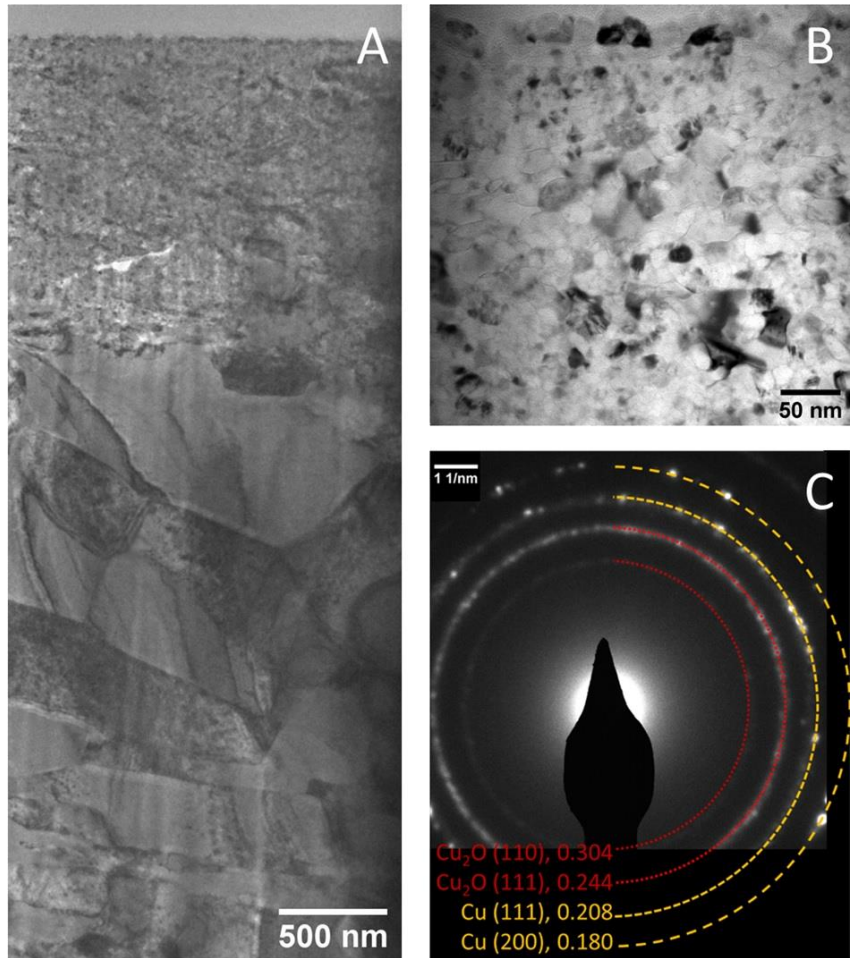

Figure 5. TEM bright field images of the grain structures formed immediately below the wear surface. A roughly 2 μm by 4 μm region below the wear surface showing the wear induced nanocrystalline region above the UFG region is given in (A). A high magnification image from the nanocrystalline region is given in (B) with the accompanying SAED pattern from this location in (C). Next to the SAED pattern are half rings showing the measured d-spacing in nm, along with the closest corresponding orientation. The sliding direction is parallel to the top of each micrograph.

## 4. Discussion
*4.1. Microstructural evolution and saturation in grain size and hardness at high equivalent strains*
Many works have characterized the microstructural evolution after HPT processing to various amounts of equivalent strain [19–23]. Dislocations introduced by HPT processing accumulate into cells within grains that increase misorientation. When sufficient dislocations accumulate after a certain level of strain, misorientation becomes large enough that these formed cells become indistinguishable from grain boundaries. Then, after sufficient straining, a saturation microstructure develops due to a balance between further dislocation generation and recovery [35]. An early work on copper HPT noted that even at high strains, grain refinement ceased below an average of about 250 nm [36]. Various works have claimed that strains between 10 and 20 are sufficient for copper to reach an equilibrium grain size [21–23,37]. From the SEM and TEM observations collected in the current work and shown in Figures 2 and 3, there are no indications that the present samples are significantly different compared to those from the literature.

While grain size may saturate after a certain amount of strain, the amount of high angle or twin boundaries may continue to evolve with further straining. Although grain size and hardness were seen to saturate at an equivalent strain of 15 in one work, it was noted that the fraction of high angle grain boundaries and average misorientation was still evolving [37]. In one EBSD investigation of pure copper HPT discs processed similarly to the current work, the mid radius of discs after 1 and 10 rotations both had similar average grain sizes and high angle grain boundary fraction [20]. Conversely, the center of the 10 rotation disc was mostly the same, but the 1 rotation disc center had larger average grain size and lower high angle grain boundary fraction. Another work characterized twin boundary fraction, which was generally low (~5%) and homogeneous in 10 rotation discs whereas the 1 rotation discs had low twinning at the center and edge but upwards of 38% in the mid-radius [38].

In addition to grain size and misorientation, defect densities also change with the amount of strain. One work measured the distribution of vacancy clusters in HPT copper discs and found that higher rotations increased vacancy concentration with larger clusters found at the outer edge compared to the center [39]. The same work also measured dislocation density and saw only a minor increase with increasing rotations. Another work used finite element methods along with original and literature XRD and TEM measurements and saw evidence that dislocation densities saturate at equivalent strains less than 10 [40]. Some works have also documented the microstructural stability of HPT copper, which is important since the discs in the current work were characterized about two years after they were initially processed. After 4 weeks, discs processed to 10 rotations had a stable dislocation density of around $1.5 \times 10^{-13}$ m$^{-2}$, while 1 rotation discs underwent recovery during that time and their dislocation density dropped by half from $6 \ 10^{-13} \times$m$^{-2}$ [41]. Thus, after 10 rotations saturation in grain size, grain boundary character, and defect density throughout the disc can be reasonably assumed, whereas a 1 rotation disc may have this saturation at the outer edge with large variation throughout.

The microhardness measurements in this study agree with several other works that have characterized copper processed by HPT and other SPD methods. Indeed, the hardening behavior of copper subjected to HPT follows from the same saturation in microstructure that was just described. HPT Cu hardness was seen to saturate around 130 HV after an equivalent strain of about 20 [42]. Copper processed by ECAE [43] and ARB [44] in the range of equivalent strains at which saturation occurs also reached around 130 HV. In the current work, the microhardness measured throughout the 10 rotation disc agrees with these values. The high microhardness in the low strain central region of the 1 rotation sample agrees with other studies that have investigated high purity copper samples aged at room temperature for times ranging from a few weeks to several years after HPT processing [38,41,45]. These works showed the occurrence of self-annealing at room temperature due to the relatively low melting temperature and stacking fault energy of high purity copper. This facilitated recrystallization in the outer edges of moderately strained discs due to higher stored energy there, while the low equivalent strain central portion remained unrecrystallized and with high hardness due to the compressive stress from the anvils. A decrease in dislocation density was seen around the mid-radius of discs processed to small (≤1) rotations, corresponding to a sharp drop in hardness before steadily increasing again. For highly strained discs (≥5 rotations), no drop in hardness was observed unless they had been aged for 7 years, after which a minor drop in the edges was found [45]. The discs in the current work were characterized two years after they were initially processed, so it is reasonable that similar recrystallization is seen in our 1 rotation sample while no appreciable change is seen in our 10 rotation sample.

Thus, at the sufficiently high equivalent strain applied to the 10 rotation sample, the hardness, average grain size, ratio of high and low grain boundaries, and defect density is thought to saturate throughout the disc. As will be seen in the following sections, this property saturation for the 10 rotation disc is reflected by a saturation in wear properties at high strains. However some differences are also seen depending on wear path, an observation which requires further exploration and consideration.

*4.2. Wear follows Archard scaling by decreasing with increasing equivalent strain and leveling off after saturation of microstructure and hardness, with a few notable exceptions*

Figure 4 shows that normalized wear rate generally decreases with increasing equivalent strain up to around 15 at which prior works have seen a saturation in microstructure and hardness. The normalized wear rate levels off at around 0.25 after this strain, i.e. only about ¼ as much wear occurs after microstructural saturation as compared to the unprocessed condition. This trend follows the commonly-cited Archard relationship between hardness and wear resistance given by Equation 2, whereby copper that is work hardened through straining will have a reduced wear rate [24]. One work on copper processed by ECAE showed a consistent drop in normalized wear rate with successive passes [46]. Based on microstructural evolution in SPD materials, this result and those from the current work are unsurprising. Since increasing strain leads to grain refinement and Hall-Petch strengthening, a sharp increase followed by a plateau in hardness should correspond with an inverse relationship for wear rate, as is illustrated in Figures 3 and 4.

However, a few works disagree with this trend between normalized wear rate and strain. Most glaring is the increased wear at high strains measured for 99.9% Cu processed by consolidating machined chips via HPT [7]. The original unprocessed coarse-grained copper conductivity in this work was only 71.7% IACS, suggesting that compositional differences were present even before any processing. It is possible that the machining and consolidation allowed further contamination from surface oxides to enter the microstructure. Embrittlement from oxides could have led to the higher wear rate measured. Another work by the same author on a Cu-0.35Cr-0.5Zr alloy also disagreed with the trends followed by the low strain works in Figure 4 [47]. In this work, increased wear and brittle fracture occurred at low strains since the elongated grains formed at this strain were easy to delaminate. Also, unlike the other alloy works in Figure 4, this composition has a supersaturation of Zr, so the zirconium left out of solution could have caused embrittlement. After 4 passes of ECAE, this zirconium may have been mechanically forced into solution giving rise to the "brittle to ductile" transition and increased wear resistance the authors observed at that processing condition. Such compositional issues encountered in these works are not well accounted for by normalizing the wear rate and make these works potentially unsuitable for comparison amongst the others.

The work on large strain extrusion machining (LSEM) in Figure 4 also shows wear behavior deviating from Archard scaling [48]. Machined chips were processed in this work at three increasing strains to create elongated nanograins, elongated/equiaxed nanograins, and equiaxed nanograins. Curiously, the greatest wear resistance was found for the lowest strain when the microstructure contained elongated nanograins. Wear rates were reported along three different sliding directions with differing results, although unfortunately these findings were not analyzed in terms of the microstructural response. In LSEM, strain is controlled by machining samples of different thicknesses, with thinner samples corresponding to higher processing strains. One possibility is that the higher wear rate seen for the high strain samples is an artifact caused by an interaction between the thinness of the sample and the experimental wear setup. It could also be possible that some aspect of the LSEM process introduced oxides or other contaminants into the microstructure. Although the authors stated that processing was slow enough to avoid heating, they do not state the speed used, and in their other works heating of 50-100°C even at low processing speeds was reported [49]. Either heating or oxidation could skew the results, or there could be more complex microstructural response as will be discussed further in the next section.

In Ni-W, another FCC system, wear has been shown to deviate from Archard scaling for grain sizes in the nanocrystalline regime [50]. Systematic investigation of pin on disc wear on nanocrystalline Ni-W showed that for the smallest grain sizes, less wear was seen than would be expected from Archard scaling. Wear rate consistently decreased with grain size, even at sizes below which the Hall-Petch relationship breaks down and hardness leveled off. This was attributed to wear induced hardening from a combination of grain growth and grain boundary relaxation. Of the works in Figure 4, the electro-deposition work and the SMGT and SMAT works had grain sizes at or past the critical size for Hall-Petch breakdown in copper. Despite these small grains, the normalized wear rates for these works arrived at the same plateau as other high strain works with no indication of the Archard deviation seen for Ni-W. Thus, it is possible that wear induced hardening for extremely small grains does not occur for copper as it did for Ni-W. The difference could lie in the two-phase Ni-W having different wear modes compared to single phase high purity copper.

In sum, this work has provided a framework by which wear behavior can be assessed at a wide range of processing strains. By focusing on characterizing normalized wear rate as a function of equivalent strain, measurements from two HPT discs were collected that generally agreed with several other works spanning multiple processing conditions. This enabled observation of a saturation in wear rate, indicating that processing to strains higher than ~20 will not improve wear behavior. Similar testing can be conducted on other materials processed by HPT to improve knowledge of processing-structure-properties relationships with regards to wear. However, the simplicity of this approach comes with a few limitations. First, normalized wear rates only go so far in negating the influence of certain variables. Large differences in solute amounts, sliding speed, or normal force can significantly alter microstructural evolution and wear behavior. Second, equivalent strain calculation is often simplified and idealized and therefore has inherent limitations in describing the microstructures that are created by SPD. At the same equivalent strain, many microstructural features like grain size, grain morphology, dislocation density, and twinning can differ for different processes, or as in the central region of HPT discs, there may be differences between the calculated and true equivalent strain experienced. Third, as with any comparative study, there are limitations based on what processing or characterization information is reported. Along with the myriad wear test parameters, a better accounting for wear track location in relation to processing textures and equivalent strains can greatly contribute to better understanding the origins of wear behavior in SPD materials.

*4.3. Wear path influences wear rates after HPT or other directional deformation processes*

The difference in wear rates between the radial and tangential tests in the current work indicates that track location and direction have a significant impact on the measured wear response for HPT discs. The higher normalized wear rates for the radial tests suggests that tracks covering a wider range of strains have a lower wear resistance. Such an observation might make sense for the inhomogeneous 1 rotation disc, but strangely it is also seen for the homogeneous 10 rotation disc. Since only one test was conducted for each location and no statistical analysis could be conducted, it is possible the difference in results between the radial and tangential tests are not truly significant. But, from other reports in the literature, there is enough indication that these results are significant enough to merit further exploration. In Figure 4, two works by Purcek et al. on Cu-0.7Cr-0.07Zr processed by HPT show a similar influence of wear path. These works used either continuous (ASTM G99) wear tests [6] at a radial distance of 5 mm or reciprocating (ASTM G133) tests [15] centered at the same radial distance. The continuous test covered a smaller range of strains and had a comparatively lower wear rate, much like what was observed in the current work. The high strains used in their studies caused deformation induced solutionization, making it unlikely that the alloying additions are the main contributors toward the wear behavior they observed.

It is possible the discrepancy between these two works by Purcek et al. is due to a difference between continuous and bidirectional sliding. The LSEM work in Figure 4 also showed differences in wear rates between continuous and reciprocating tests, although without clear trends. Multiple prior works on coarse-grained aluminum, copper, and steel alloys have seen higher wear rates for unidirectional wear [51–53]. Some of these works attributed their results to the Bauschinger effect, by which a reversal in direction of an applied stress is accompanied by a decrease in yield strength. Strain hardening defects generated from stress in one direction may be annihilated upon the change in direction, resulting in softening. While these works did see a decrease in hardness after a few changes in direction in agreement with the Bauschinger effect, this softening did not manifest in an increase in wear rate as might be expected. The microstructures from HPT may have a complex interaction with the Bauschinger effect, thereby giving rise to the wear response seen in the current results and the two works by Purcek et al., but validating this would require further investigation. As a starting point, one computational study found the number of defects retained is higher in unidirectional sliding of single crystalline and nano-twinned microstructures, but not for nanograined copper [54]. Thus, it is possible that changes in grain boundary character (i.e. twinning amount) for strained samples can be contributing to their wear response during continuous tests. This may also explain the unexpected trend between strain and wear rate for the LSEM work. Since the Bauschinger effect is essentially a microstructural response, it is vital to further discuss how wear behavior changes due to differences in microstructures encountered along wear tracks.

The influence of crystallographic orientation on wear response can be inferred from some of the data in Figure 4, and it has also been described in works on other materials over the years. In the current work, due to the shear texture that forms in HPT discs, the tangential tests with lower wear rate are more parallel to the shearing direction, whereas the radial tests with higher wear rate are entirely perpendicular to the torsional shearing. One work from NASA in 1969 reported that 50% rolled aluminum displayed lower wear when slow, mild sliding was normal to the rolling direction compared to parallel [55]. The work on extruded copper included in Figure 4 also showed decreased wear rate when disc on shoe sliding was parallel to the extrusion direction with an increase in wear rate for perpendicular sliding [56]. A work on an aluminum alloy processed by ECAP also had different wear rates when parallel or normal to the deformation axis [57]. In a work with polycrystalline cold rolled/drawn steel pins worn on steel discs, changes in wear properties were observed depending on the orientation with which the pin was cut, and a higher wear rate was seen when sliding was parallel to the rolling or drawing direction [58]. Another work on single crystals of copper noted differences in wear particle formation depending on the orientation of the shear plane in relation to the sliding [59]. The influence of texture on wear and vice versa has been studied in a series of works using bronze alloy pins in continuous pin on disc wear tests by Cai et al. Recently they quantitatively described strain accumulation and grain rotation for single crystal and coarse grained polycrystalline pins [60]. All of these results point toward the existence of a unified explanation for the crystallographic influence on wear properties, one area that tribologists in the SPD community are uniquely equipped to investigate.

While the current work has focused on post mortem determination of wear rates after dry sliding wear, another related work monitored changes in deformation behavior due to lubricated sliding on metals processed at different initial strains [61]. This study used *in-situ* video of a lubricated wedge sliding across aluminum samples at initial strains from 0 to 2 to determine the mesoscale strain accumulation after a single pass. The authors characterized the microstructural response in the wake of this sliding as "laminar" for these low strains, and higher initial strains incurred a breakdown in this smooth response, with folds or localized shear bands created instead. Formation of such inhomogeneities leads to material removal and increases wear rate. Another computational work eschewed the flow-like treatment of fold formation, and instead presented evidence that bulging is a crystallographic response by grains of ideal orientations with respect to the sliding force field [62]. It is possible that the shear textures imposed by HPT processing may be better oriented for bulge formation when sliding in certain directions, but more work needs to be done to verify this.

Outside of the SPD community, the work done by Greiner et al. has illuminated the microstructural response during early stages of dry sliding for coarse grained copper [32]. The coefficients of friction in that work are in the same range as those in the current work, further validating that mild sliding was achieved despite different test parameters. One important feature of this work is that it has shown that dislocations injected from the wear surface accumulate in the microstructure due to sliding, forming a "dislocation trace line" a certain depth below the surface. A better understanding of this dislocation trace line formation in relation to any potential Bauschinger effects is needed. The formation of amorphous oxygenated clusters at the wear surface in early sliding was also reported in this work, observing a pathway by which oxygen enters the nanocrystalline layer, as was observed from the SAED in the current work. Recently, the same research group has reported another dry sliding study using coarse grained copper membranes with different aspect ratios to investigate differences in tribology with strain [63]. This membrane work analyzed strain during sliding as opposed to the current work which analyzed initial processing strain. They found that membranes with higher aspect ratios experienced more strain during sliding and exhibited lower COF and less geometrically necessary dislocations in the subsurface microstructure. Insight gained on the fundamental processes in wear of coarse-grained materials can be applied toward understanding wear of fine-grained materials and vice versa to improve basic understanding of wear mechanisms across all grain sizes.

To summarize, researchers should be aware that different test locations can give different measurements of wear response in HPT discs. This should be kept in mind when designing wear studies, reporting results, and comparing the results of others. Prior works on SMAT and other surface deformation techniques have probed the wear response of materials with large gradients in equivalent strain and grain size below the tracks. This work is unique in analyzing the wear response of samples with a variety of such gradients along the length of the track. Since HPT produces materials with substantial radial inhomogeneity, they are prime candidates for exploring tribological response across nonuniform microstructural regimes. Further research can more comprehensively probe the microstructural response when sliding wear traverses a wide range of grain sizes.

## 5. Conclusion

This work has examined variations in wear rate with respect to equivalent strain in HPT processed copper discs. Microstructural saturation after processing past an equivalent strain of about 15 reduced wear rate by about 75% compared to unprocessed copper. Additionally, discrepancies between linear reciprocating wear tests conducted at different regions of the same disc has indicated that wear resistance is dependent on a relationship between strain path and wear path. Wear measurements gathered only from 1 and 10 rotation HPT discs agreed with values from the literature for various SPD conditions. The incremental microstructural and wear evolution shown for high equivalent strains indicate that materials receive no tribological benefits from higher SPD strains.

**Acknowledgements**

HPT was carried out in the International Research Center on Giant Straining for Advanced Materials (IRC-GSAM) at Kyushu University, Japan. Scanning electron microscopy was performed on a FEI Nova NanoSEM 450 in the Central Facility for Advanced Microscopy and Microanalysis (CFAMM) at UC Riverside. Focused ion beam was performed on a ThermoFisher Scientific (formerly FEI/Philips) Quanta™ 3D 200i in the CFAMM at UC Riverside. Transmission electron microscopy was performed on a ThermoFisher Scientific (formely FEI/Philips) Titan Themis 300 and an FEI Tecnai T12 in the CFAMM at UC Riverside. EHR was supported by the GAANN Fellowship from the Department of Education through the Mechanical Engineering department at the University of California, Riverside. SNM was supported via NSF CMMI Grant # 1663522.